\documentclass[twocolumn,preprintnumbers,amsmath,amssymb,aps,showkeys]{revtex4}
\usepackage{graphicx}
\usepackage{dcolumn}
\usepackage{bm}
\usepackage{float}
\usepackage{color}
\usepackage{soul}
\usepackage{mathtools}
\usepackage{changes}
\setdeletedmarkup{\textcolor{red}{\sout{#1}}}   
\setaddedmarkup{\textcolor{blue}{#1}} 
 
\begin{document}
	
	\title{Dynamical Quantum Phase Transitions in Periodically Driven 1D Ising Model}
		
	\author{Yuanyuan Cheng}
	\affiliation{School of Science, Qingdao University of Technology, Qingdao 266525, China.}	
	
	\author{Yuxia Zhang}
	\email{zhangyuxia0619@163.com}
	\affiliation{School of Science, Qingdao University of Technology, Qingdao 266525, China.}
	
	\author{Tianhui Qiu}
	\affiliation{School of Science, Qingdao University of Technology, Qingdao 266525, China.}
	
	\author{Peipei Xin}
	\affiliation{School of Science, Qingdao University of Technology, Qingdao 266525, China.}
	
	\author{Bao-Ming Xu}
	\email{xubm2018@163.com}
	\affiliation{Institute of Biophysics, Dezhou University, Dezhou 253023, China}

	\date{Submitted \today}

	\begin{abstract}
		This work investigates dynamical quantum phase transitions (DQPTs) in a one-dimensional Ising model subjected to a periodically modulated transverse field. In contrast to sudden quenches, we demonstrate that a DQPT can be induced in two distinct ways. First, when the system remains within a given phase--ferromagnetic (FM) or paramagnetic (PM), a resonant periodic drive can trigger a DQPTs when its frequency matches the energy-level transition of the system. This DQPT is intimately connected to the emergence of Floquet topological phases. The timescale for the transition is governed by the perturbation strength $\lambda'$, the critical mode $k_c$, and its energy gap $\Delta_{k_c}$, following the scaling relation $\tau\propto\Delta_{k_c}\lambda'^{-1}\csc k_c$. Second, for drives across the critical point between the FM and PM phases, low frequencies can always induce DQPT, regardless of resonance. This behavior stems from the degeneracy of the energy-level at the critical point, which ensures that any drive with a frequency lower than the system’s intrinsic transition frequency will inevitably excite the system. However, in the high-frequency regime, such excitation will be strongly suppressed, thereby inhibiting the occurrence of DQPTs. This study provides deeper insight into the nonequilibrium dynamics of quantum spin chains.
	\end{abstract}
	    
 	\maketitle
	\section{Introduction}
	With these experimental advances for exploring the non-equilibrium dynamics of confined quantum systems such as ultracold atoms \cite{Schreiber2015,Jotzu2014,Daley2012}, Rydberg atoms \cite{Saffman2010}, superconducting qubits \cite{Klocke2025}, and ion traps \cite{Cormick2010,Knaute2022}, an in-depth study of the dynamics of quantum many-body systems far from thermodynamic equilibrium becomes possible. Dynamical quantum phase transitions (DQPTs) have recently
	emerged as an interesting phenomenon within this regime \cite{Heyl2015,Heyl2013,Polkovnikov2011,RevModPhys.83.863,Zurek2005,Heyl2018,Heyl2019}, which serve as a nonequilibrium counterpart to equilibrium phase transitions. The concept of DQPTs arises from the analogy between the equilibrium partition function of a system and the Loschmidt amplitude, which measures the overlap between an initial state and its time-evolved state \cite{Heyl2015, Quan2006,Jafari2019a,Jafari2017,Divakaran2013,Hou2020,Najafi2019,Yan2020,Najafi2018,Mukherjee2019}. As equilibrium phase transitions are signaled by non-analyticities in thermal free energy, DQPTs are revealed through nonanalytical behavior in dynamical free energy \cite{Andraschko2014,Karrasch2013,Vajna2014,Jafari2021,Vajna2015,Mendoza-Arenas2022,Mondal2022,Yan2020,DeNicola2021,Sedlmayr2018a,Sedlmayr2018b,Verga2023,Khan2023}, with real time serving as the control parameter \cite{Heyl2019,Karrasch2013,Kriel2014,Canovi2014,Calabrese2006,Uhrich2020,DeNicola2021,Yu2021,Sadrzadeh2021,Liou2018}. The non-analyticity in quantum state evolution is accompanied by universal critical exponents, unveiling scaling behavior and universality class patterns near quantum critical times \cite{Bhattacharjee2018,Heyl2015,Yao2025,Trapin2021}. Furthermore, analogous to order parameters in equilibrium quantum phase transitions, a dynamical topological order parameter has been proposed to capture DQPTs \cite{Bhattacharjee2018, Sharma2014,Bhattacharya2017a,Eckstein2009}. Experimentally, DQPT has been realized on various quantum simulators, such as ultracold atoms \cite{Flaeschner2018}, trapped ions \cite{Jurcevic2017}, superconducting quantum circuits \cite{Guo2019}, etc \cite{Chen2020a,Wang2019,Xu2020,Zhang2025}.
	
	DQPTs was originally proposed in the context of quench dynamics within a nearest-neighbor transverse field Ising model \cite{Heyl2013,Heyl2015,Bhattacharjee2018,PhysRevB.107.L121113}, where the transverse field of the system Hamiltonian is suddenly changed. It was later extended to finite-time slow quench processes \cite{DeGrandi2010,Weiss2018,Sim2022,Cao2024,Poletti2011,Chen2020b}. Due to the critical slowing-down effect near a quantum phase transition, crossing the critical point at any finite rate inevitably excites the system, thereby making quantum coherence an essential feature of such transitions-a behavior captured by the Kibble-Zurek mechanism \cite{DeGrandi2010,Weiss2018,Sim2022,Kang2020,Lee2015,Balducci2023}. Notably, quantum coherence not only restores DQPTs that would otherwise be destroyed by thermal fluctuations, but can also give rise to entirely new DQPTs that are independent of the equilibrium quantum critical point \cite{Xu_2024}. In recent years, periodically driven systems have opened a new avenue for exploring DQPTs \cite{Quelle2017, Ansari2025, Kitagawa2010, Yang2021, Brown2021, Kriel2014, Bordia2017, Murali2025, Nandy2018}. Unlike conventional quench dynamics, periodic driving allows Hamiltonian parameters to be modulated in diverse ways \cite{Quelle2017}, leading to a wide range of experimental schemes and platforms for realizing driven settings. Although numerous studies have examined DQPTs from various perspectives, a comprehensive understanding of their underlying mechanisms remains elusive.
	
	To address this issue, this work studies DQPTs in a one-dimensional (1D) Ising model driven by a periodic transverse field. Using time-dependent perturbation theory, we examine the roles of the driving intensity and frequency, which elucidates the underlying mechanism of DQPTs. Our results show that the driving frequency is crucial for the occurrence of a DQPTs. Specifically, a DQPT arises when the driving frequency resonates with the energy-level transition frequency of the system. The driving strength, together with the critical mode and its energy gap, governs the timescale required for the DQPTs to occur, satisfying the scaling law $\tau\propto\Delta_{k_c}\lambda'^{-1}\csc k_c$. Furthermore, we show that low-frequency perturbations that cross the equilibrium critical point can also induce DQPTs, whereas high-frequency drives strongly suppress the evolution of the system.
	
	The structure of this paper is as follows:
	Sec. \ref{Sec: Model and Periodic Driving Protocol} introduces the periodically driven Ising model and reviews the basic concepts of DQPT. In Sec. \ref{Sec: The Influence of periodic driving Within the Same phase}, we elucidate the roles of frequency and strength of the drive on DQPT when the system remains within the same equilibrium phase. Sec. \ref{Sec: The Influence of periodic driving across the critical point} investigates the influences of periodic driving on DQPTs when the system crosses the equilibrium quantum critical point. Finally, Sec. \ref{Sec: Conclusion} closes the paper with some concluding remarks.

	\section{Model and Periodic Driving Protocol}\label{Sec: Model and Periodic Driving Protocol}
	The Hamiltonian of the 1D transverse field Ising model is
	\begin{equation}\label{}
		\hat{H}=-\frac{J}{2}\sum_{j=1}^{N}\bigl[\hat{\sigma}_j^z\hat{\sigma}_{j+1}^z+\lambda\hat{\sigma}_j^x\bigr],
	\end{equation}
	where $\hat{\sigma}^{\alpha}_{j}$ $(\alpha=x,y,z)$ is the spin-1/2 Pauli operator at lattice site $j$ and the periodic boundary condition is imposed as $\hat{\sigma}^{\alpha}_{N+1}=\hat{\sigma}^{\alpha}_{1}$. Here we only consider that $N$ is even. $J$ is longitudinal spin-spin coupling, and we set $J=1$ as the overall energy scale without loss of generality. $\lambda$ is a dimensionless parameter measuring the strength of the transverse field with respect to the longitudinal spin-spin coupling. Driven by the transverse field $\lambda$, a quantum phase transition from the ferromagnetic (FM) phase ($\lambda<1$) to the paramagnetic (PM) phase ($\lambda>1$), which is known as the Ising transition, occurs at the critical point of $\lambda_{c}=1$.
	
	This Hamiltonian is integrable and can be mapped to a system of free fermions and therefore be solved exactly. By applying the Jordan-Wigner transformation and the Fourier transformation, the Hamiltonian converts from spin operators into spinless fermionic operators as \cite{Puskarov2016}
	\begin{equation}\label{Hamiltonian}
		\hat{H}=\sum_{k>0}
		\begin{pmatrix} \hat{c}^\dag_{k} & \hat{c}_{-k} \end{pmatrix}
		\begin{pmatrix} \lambda-\cos k &-i\sin k \\ i\sin k & -\lambda+\cos k\end{pmatrix}
		\begin{pmatrix} \hat{c}_{k} \\ \hat{c}^\dag_{-k} \end{pmatrix},
	\end{equation}
	where $\hat{c}_{k}$ and $\hat{c}^\dag_{k}$ are respectively fermion annihilation and creation operators for mode $k=(2n-1)\pi/N$ with $n=1,2,\cdots, N/2$, corresponding to antiperiodic boundary conditions when $N$ is even. Each $\hat{H}_k$ acts on a two-dimensional Hilbert space generated by $\{\hat{c}^\dag_{k}\hat{c}^\dag_{-k}|0\rangle,~|0\rangle\}$, where $|0\rangle$ is the vacuum of the Jordan-Wigner fermions $\hat{c}_k$ and $\hat{c}_{-k}$, and can be represented in that basis by a Pauli matrix:
    \begin{equation}\label{}
        \hat{H}_k=(\lambda-\cos k)\hat{\sigma}^z+\sin k\hat{\sigma}^y.
    \end{equation}
    
    The energy levels of $\hat{H}_k$ are
	\begin{equation}\label{ground energy}
		\varepsilon^{\pm}_k=\pm\varepsilon_k=\pm\sqrt{(\lambda-\cos k)^2+\sin^2 k},
	\end{equation}
	with the corresponding eigenvectors
	\begin{equation}\label{}
		\begin{split}
			|\varepsilon^{+}_{k}\rangle&=\biggl[i\sin\frac{\theta_k}{2}+\cos\frac{\theta_k}{2}\hat{c}^\dag_{k}\hat{c}^\dag_{-k}\biggr]|0\rangle, \\
			|\varepsilon^{-}_{k}\rangle&=\biggl[\cos\frac{\theta_k}{2}+i\sin\frac{\theta_k}{2}\hat{c}^\dag_{k}\hat{c}^\dag_{-k}\biggr]|0\rangle,
		\end{split}
	\end{equation}
	where $|0\rangle$ is the vacuum state. The angle $\theta_k$, which is generally complex, is determined by $\tan\theta_k=\sin k/(\lambda-\cos k)$.
	
	At time $t<0$, the system is prepared in the ground state
	\begin{equation}\label{}
		|G\rangle=\bigotimes_{k>0}|\varepsilon^{-}_{k}\rangle.
	\end{equation}
	A periodic perturbation is then switched on at $t=0$, characterized by the transverse field
	\begin{equation}\label{instantaneous field}
		\lambda(t)=\lambda+\lambda'\cos(\omega t),
	\end{equation}
	with amplitude $\lambda'$ and frequency $\omega$. In the following, we will consider three types of driving protocols based on the instantaneous value of $\lambda(t)$ relative to the static critical point $\lambda_c = 1$: (i) $\lambda(t) < 1$ for all $t$, i.e., $\lambda + \lambda' < 1$, where the instantaneous ground state remains entirely in the FM phase; (ii) $\lambda(t) > 1$ for all $t$, i.e., $\lambda - \lambda' > 1$, where the instantaneous ground state remains entirely in the PM phase; (iii) $\lambda(t)$ crosses $\lambda_c = 1$, i.e., $\lambda - \lambda' < 1 < \lambda + \lambda'$, where the instantaneous ground state undergoes a phase transition. For convenience, we label these protocols by the corresponding static phase(s) involved. It is worth noting that periodically driven systems can exhibit Floquet or topological phase transitions, which are governed by the quasienergy spectrum of the Floquet Hamiltonian \cite{Kitagawa2010,Bukov2015,Rudner2013,Lindner2011}. The associated phase diagrams are typically parameterized by driving parameters such as frequency and amplitude. Nevertheless, the present work focuses on the real-time dynamics of the system and aims to investigate the emergence of DQPTs during the driving process.
    
	The dynamics of the system is governed by the perturbed Hamiltonian $\hat{H}(t)$, and the system state at time $t$ is determined by ($\hbar=1$)
	\begin{equation}\label{}
		|\psi(t)\rangle=\mathcal{T}\exp\Bigl[-i\int_0^t\hat{H}(t')dt'\Bigr]|G\rangle,
	\end{equation}
	where $\mathcal{T}$ denotes the time-ordering operator. The instantaneous state can be expressed in a tensor product form as $|\psi(t)\rangle=\bigotimes_{k>0}|\psi_k(t)\rangle$, where each mode evolves as
	\begin{equation}\label{}
		|\psi_k(t)\rangle=(v_{k}(t)+ u_{k}(t)c_k^\dagger c_{-k}^\dagger)|0\rangle.
	\end{equation}
	The coefficients $u_{k}(t)$ and $v_{k}(t)$ can be obtained by solving the differential equation
	\begin{equation}\label{equation}
		\frac{d}{dt}\begin{pmatrix} u_{k}(t) \\ v_{k}(t) \end{pmatrix}
		=-i\begin{pmatrix} \lambda(t)-\cos k &-i\sin k \\ i\sin k & -\lambda(t)+\cos k\end{pmatrix}
		\begin{pmatrix} u_{k}(t) \\ v_{k}(t) \end{pmatrix}.
	\end{equation}
	
	The main goal of this paper is to investigate DQPT induced by periodic driving.
	The theory of DQPT is built upon the Loschmidt overlap amplitude,
	\begin{equation}\label{}
		\mathcal{G}(t) = \langle G|\psi(t)\rangle=\prod_{k>0}\langle\varepsilon^{-}_{k}|\psi_k(t)\rangle,
	\end{equation}
	which measures the overlap between the time-evolved and initial states. A DQPT is signaled by the vanishing of this amplitude, indicating orthogonality between $|\psi(t)\rangle$ and $|G\rangle$. This criticality is identified by a non-analytic cusp in the dynamical free energy density, or the rate function,
	\begin{equation}\label{}
		r(t) = -\frac{1}{\pi} \int_0^\pi dk \ln|\langle\varepsilon^{-}_{k}|\psi_k(t)\rangle|^2,
	\end{equation}
	serving as the counterpart of the free energy in equilibrium phase transitions.
	Moreover, DQPTs are endowed with topological properties, characterized by the winding number defined as \cite{Bhattacharya2017b}
	\begin{equation}
		\nu(t) = \frac{1}{2\pi} \int_0^\pi \frac{\partial \phi_k^G(t)}{\partial k} dk,
	\end{equation}
	which undergoes a quantized jump at the critical times of the DQPT. The underlying geometric phase is defined as
	\begin{equation}\label{}
		\phi_k^G(t)=\phi_k(t)-\phi_k^D(t),
	\end{equation}
	where
	\begin{equation}\label{}
		\phi_k(t)=\arg\mathcal{G}_k(t)
	\end{equation}
	is the total phase, and
	\begin{equation}\label{}
		\phi_k^D(t)=-\int_0^t\langle\psi_k(t')|\hat{H}_k(t')|\psi_{k}(t')\rangle dt'.
	\end{equation}
	is the dynamical phase.
		
	\begin{figure}
		\begin{center}
			\includegraphics[width=8cm]{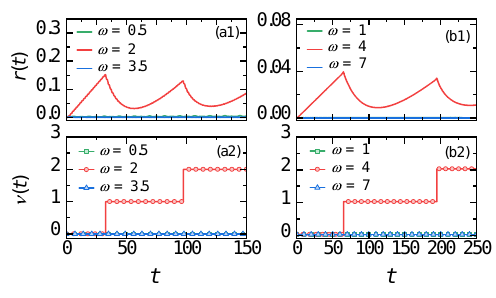}
			\caption{(Color online) Time evolution of the rate function $r(t)$ and the winding number $\nu(t)$ under different driving frequencies. Panels (a1-a2) depicts the magnetic field initially fixed within the FM phase $\lambda = 0.5$. Panels (b1-b2) depicts the magnetic field initially fixed within the PM phase $\lambda = 2$. The driving strength is $\lambda'=0.1$ for all the cases.}
			\label{fig1}
		\end{center}
	\end{figure}

    Due to the time-dependent nature of the Hamiltonian, an analytical solution to Eq.~\eqref{equation} is not feasible. We therefore solve the equation numerically to obtain exact results, which serve as the basis for investigating DQPTs in the periodically driven Ising model. To gain deeper insight into the underlying mechanisms of these dynamical phase transitions, we complement the numerical approach with a time-dependent perturbation analysis. 
	
	\section{The Influence of periodic driving Within the Same phase}
	\label{Sec: The Influence of periodic driving Within the Same phase}
	
	In this section, we explore the emergence of DQPTs driven by periodic driving within the FM phase ($\lambda+\lambda'<1$) or PM phase ($\lambda-\lambda'>1$). First, we investigate the effects of the driving frequency for a given magnitude $\lambda'=0.1$. Fig. \ref{fig1} illustrates the time evolution of the Loschmidt echo rate function and the winding number for different frequencies. The results demonstrate a critical dependence of DQPTs on the driving frequency. Pronounced non-analytic signatures are observed only when the driving frequency $\omega$ is tuned to a specific regime, for instance, $\omega=2$ for $\lambda=0.5$ (FM phase) or $\omega=4$ for $\lambda=2$ (PM phase). Under these resonant conditions, the rate function exhibits clear oscillatory behavior with periodic cusp singularities. Concurrently, the winding number, which remains quantized during the dynamics, undergoes discrete integer jumps precisely at the critical times where the cusps appear. The simultaneous occurrence of these features provides definitive evidence for a DQPT. By contrast, when the driving frequency is detuned from these specific values, the rate function stays close to zero throughout the evolution, indicating that the system does not evolve significantly from its initial state, and no DQPT occurs. These observations underscore that the resonant frequency plays a crucial role in the occurrence of DQPT.
    
	\begin{figure}
		\begin{center}
			\includegraphics[width=8cm]{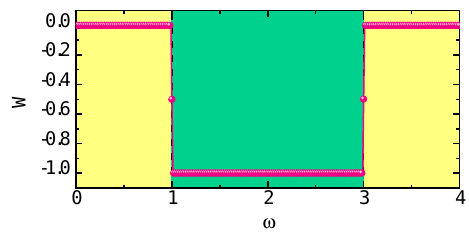}
			\caption{(Color online) The winding number $W$ as a function of $\omega$ for $\lambda=0.5$ and $\lambda'=0.1$.}.
			\label{fig2}
		\end{center}
	\end{figure}
    
	A comparison between the results of periodic driving and sudden quench dynamics is essential. Although a sudden quench performed within a given phase of the Ising model cannot induce a DQPT \cite{Cao2022,Puskarov2016}, we demonstrate here that a periodic perturbation applied within the same phase can indeed trigger a DQPT, provided that the driving frequency is tuned into an appropriate range. This crucial frequency dependence originates from transitions among the system's energy levels, revealing the physical mechanism behind DQPT.
    
	As clarified by time-dependent perturbation theory (see Appendix \ref{Sec: A} for details), a periodic perturbation can efficiently drive transitions between energy levels only when its frequency resonates with the intrinsic energy gap of the system. In contrast, off-resonant perturbations have negligible effects, leaving the system essentially unchanged. For example, to induce transitions between the energy levels of the $k$ mode, whose energy gap is $\Delta_k=2\varepsilon_k$, the driving frequency must satisfy $\omega=2\varepsilon_k$. Such resonant driving is necessary for the system to evolve far from its initial state and undergo a DQPT. From the energy dispersion $\varepsilon_k =\sqrt{\lambda^2 - 2\lambda \cos k + 1}$, the energy gap $\Delta_k=2\varepsilon_k$ varies over the interval $2|\lambda-1| < \Delta_k < 2|\lambda+1|$ as $k$ goes from $0$ to $\pi$. Therefore, whenever the driving frequency lies within this interval, it can always resonate with the energy gap of some mode--specifically, the critical mode $k_c = \arccos\bigl(\frac{\lambda^2 + 1 - \omega^2/4}{2\lambda}\bigr)$--thereby inducing a DQPT. We refer to the frequency range of $2|\lambda-1| < \omega < 2|\lambda+1|$ as the resonant region. It should be noted that perturbations with the boundary frequencies $2|\lambda-1|$ and $2|\lambda+1|$ do not induce a DQPT, because the corresponding critical modes are $\pi$ and $0$, respectively. In these cases, the Hamiltonians before and after the perturbation commute, so the initial ground state remains unchanged and does not evolve, let alone undergo a DQPTs.

	\begin{figure}
		\begin{center}
			\includegraphics[width=8cm]{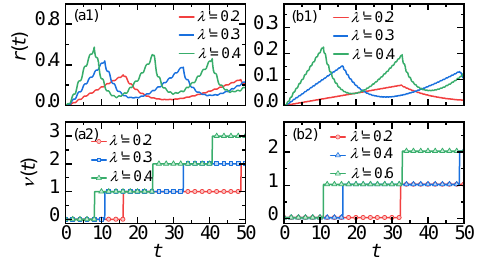}
			\caption{(Color online) Time evolution of the rate function $r(t)$ and the winding number $\nu(t)$ under different disturbance intensities. Panels (a1-a2) depicts the evolutionary process within the FM phase, where $\lambda = 0.5$, $\omega = 2$. Panels (b1-b2) depicts the evolutionary process within the PM phase, where $\lambda = 2$, $\omega = 4$.}
			\label{fig3}
		\end{center}
	\end{figure}

    Interestingly, this DQPT is governed by a deeper topological mechanism. To reveal this mechanism, we focus on the Floquet Hamiltonian:
    \begin{equation}\label{}
        \hat{H}_k^{F} = -\frac{\delta_k}{2} \tilde\sigma_k^z - \frac{\lambda'\sin\theta_k}{2} \tilde\sigma_k^y, 
    \end{equation}
    where $\tilde\sigma_k^z = |\varepsilon_k^+\rangle\langle\varepsilon_k^+| - |\varepsilon_k^-\rangle\langle\varepsilon_k^-|$, $\tilde\sigma_k^x = |\varepsilon_k^+\rangle\langle\varepsilon_k^-| + |\varepsilon_k^-\rangle\langle\varepsilon_k^+|$, and 
    $\tilde\sigma_k^y = -i|\varepsilon_k^+\rangle\langle\varepsilon_k^-| + i|\varepsilon_k^-\rangle\langle\varepsilon_k^+|$. Here $\delta_k = \omega - 2\varepsilon_k$ denotes the detuning of the driving frequency. A detailed derivation is provided in Appendix~\ref{Sec: B}. This Hamiltonian defines a vector $\textbf{d}(k) = \bigl(0, -\frac{\lambda'\sin\theta_k}{2}, -\frac{\delta_k}{2}\bigr)$, whose trajectory as a function of $k$ forms a closed curve confined to the $y$-$z$ plane. The number of times this vector encircles the origin of the $y$-$z$ plane as $k$ varies from $-\pi$ to $\pi$ defines a winding number:
    \begin{equation}\label{}
        W = -\frac{1}{2\pi i} \int_{-\pi}^{\pi} \frac{\partial}{\partial k} \arg \textbf{r}(k) ~ dk, 
    \end{equation}
    where $\textbf{r}(k) = d_y(k) + i d_z(k)$. In Fig.~\ref{fig2}, we plot the winding number as a function of $\omega$. Strikingly, we find that when the driving frequency enters the resonant region $2|\lambda-1| < \omega < 2|\lambda+1|$---where DQPTs occur---the system is driven into a topologically nontrivial Floquet phase. In other words, the DQPTs considered here are triggered by Floquet topological phases: DQPTs occur if and only if the effective Floquet Hamiltonian $\hat{H}_k^{F}$ is topologically nontrivial, and vice versa. The resonance condition is fundamentally equivalent to the boundary condition for entering a nontrivial topological Floquet phase, and the observed Rabi-flipping DQPT constitutes the dynamical manifestation of this underlying topological structure. This finding is fully consistent with the results of Ref.~\cite{Yang2019}, where such behavior was identified and termed a Floquet DQPT.
    
	Next, we examine how the strength of the perturbation influences DQPTs. Fig. \ref{fig3} shows the effect of varying the perturbation strength on the first critical time at which DQPTs occurs for a fixed resonant frequency. It is observed that as the strength $\lambda'$ increases, the first critical time becomes shorter. This trend holds consistently in both the FM and PM phases. We interpret this first critical time as the characteristic time $\tau$ required to induce the DQPT.
	
	\begin{figure}
		\begin{center}
			\includegraphics[width=8cm]{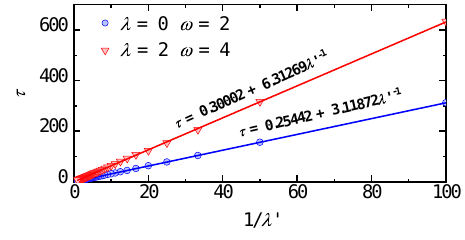}
			\caption{(Color online) The scaling relation between the DQPT onset time $\tau$ and the driving strength $\lambda'$ under resonant driving.}.
			\label{fig4}
		\end{center}
	\end{figure}
	
	To establish a quantitative relationship between driving strength $\lambda'$ and DQPT onset time $\tau$, we analyze their scaling behavior. Fig. \ref{fig4} shows the dependence of $\tau$ on the perturbation strength $\lambda'$ at a fixed resonant frequency, with the system initialized in the FM phase ($\lambda=0$, $\omega=2$) and the PM phase ($\lambda=2$, $\omega=4$). In our numerical calculations, $\lambda'$ is restricted to the interval $0<\lambda'<1$, which ensures $0+\lambda'<1$ (FM phase) and $2-\lambda'>1$ (PM phase), so that the system remains entirely within the corresponding phase. In both cases, the time required to induce a DQPT scales inversely with the perturbation strength, following $\tau \sim c \lambda'^{-1}$. This inverse scaling can be understood within perturbation theory (see Appendix for details). The amplitude of the excited state evolves as $a_k^{+}(t) \approx \lambda' \sin \theta_k t$ (see Appendix for details), implying that the characteristic time for exciting the system scales as $t \propto (\lambda' |\sin \theta_{k}|)^{-1}$. This excitation causes the return probability to the ground state to vanish, thereby inducing a DQPT. Since the DQPTs is governed by the critical mode $k_c$, the onset time $\tau$ corresponds precisely to the energy-level-transition time at $k_c$: $\tau \propto (\lambda' |\sin \theta_{k_c}|)^{-1}$. Using the relation $\sin \theta_{k} = \sin k / \varepsilon_k = 2 \sin k / \Delta_k$, we derive the scaling law for the DQPT onset time:
	\begin{equation}\label{scaling}
		\tau\propto\Delta_{k_c}\lambda'^{-1}\csc k_c,
	\end{equation}
	where $\Delta_{k_c}$ denotes the energy gap corresponding to the critical mode $k_c$. This relationship indicates that a larger energy gap results in a longer DQPT onset time $\tau$, since it directly extends the duration needed for the system to undergo the underlying energy-level transition. Remarkably, an arbitrarily weak perturbation is sufficient to induce a DQPT, provided the evolution time is sufficiently long. However, if the critical mode is $k_c = 0$ or $\pi$, the onset time $\tau$ diverges, and consequently no DQPT occurs--consistent with our earlier conclusion. Conversely, the onset time is minimized when the critical mode is $k_c = \pi/2$.
	
	For the representative FM case $(\lambda = 0, \omega = 2)$, all modes except $k=0$ and $k=\pi$ become critical that are capable of inducing DQPTs. Among these, the critical mode $k_c = \pi/2$ yields the shortest onset time: $\tau_1 \propto 2\lambda'^{-1}$. In contrast, for the PM phase $(\lambda = 2, \omega = 4)$, the critical mode is $k_c = \arccos(1/4)$, leading to an onset time: $\tau_2 \propto 16/(\sqrt{15}\,\lambda')$. Consequently, the time required to trigger a DQPT in the PM phase is approximately twice that in the FM phase, as illustrated in figure \ref{fig4}. This observation provides clear confirmation of the scaling law presented in equation \eqref{scaling}.
	
	\begin{figure}
		\begin{center}
			\includegraphics[width=8cm]{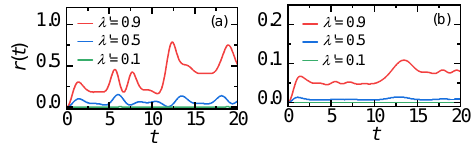}
			\caption{(Color online) The time evolution of the rate function in the (a) FM ($\lambda=0$) and (b) PM ($\lambda=2$) phases for different driving intensities. The frequency is $\omega=0.5$ for all cases.}
			\label{fig5}
		\end{center}
	\end{figure}
	
	A natural question is whether increasing the intensity of the detuning perturbation can induce a DQPT. To address this, we calculate the rate function for different perturbation intensities, as shown in Fig. \ref{fig5}. It can be seen that the rate function increases with the perturbation strength in both the FM and PM phases, but the cusp singularity cannot be observed, indicating that DQPTs cannot be induced solely by enhancing the detuning perturbation.
	
	\section{The Influence of periodic driving across the critical point}
	\label{Sec: The Influence of periodic driving across the critical point}
	
\begin{figure}
		\begin{center}
			\includegraphics[width=8cm]{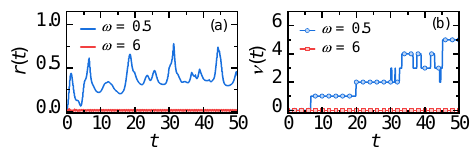}
			\caption{(Color online) The time evolution of the rate function (a) and the winding number (b) at different drive frequencies when the transverse magnetic field crosses the quantum critical point, where $\lambda = 0.5$ and $\lambda'=2$.}
			\label{fig6}
		\end{center}
	\end{figure}

	In this section, we examine perturbations that drive the system across the critical point at $\lambda_c = 1$, specifically in the regime where $\lambda < 1$ and $\lambda + \lambda' > 1$. Since Fig. \ref{fig1} established that any driving with frequency within the resonance interval $2|\lambda-1| < \omega < 2|\lambda+1|$ invariably induces a DQPT, we focus here on off-resonant frequencies outside this range. Fig. \ref{fig6} shows the time evolution of the Loschmidt echo rate function and the winding number for the protocol $\lambda(t) = 0.5 + 2\cos(\omega t) $, with $\omega = 0.5$ (low frequency) and $\omega = 6$ (high frequency). The results demonstrate that low-frequency interphase perturbations can indeed induce DQPTs, as signaled by the characteristic cusp singularities in the rate function and discrete jumps in the winding number. In contrast, under the influence of high-frequency interphase perturbations, the system does not evolve significantly, let alone DQPT.

	It is well known that an interphase quench, which corresponds to the limiting case of $\omega = 0$ in our setup, guarantees a DQPT \cite{Cao2024,Bhattacharya2017a,Puskarov2016}. In this light, our findings indicate that, while low-frequency perturbations maintain this behavior, high-frequency driving can effectively suppress the occurrence of DQPTs. This distinction can be understood from the following physical picture. Under low-frequency driving, the system evolves slowly enough to respond to the changing Hamiltonian. Although the driving frequency lies outside the resonant regime, crossing the critical point $\lambda_c = 1$ inevitably excites the system. At the critical point, the ground state becomes degenerate and the energy gap closes. According to the Kibble‑Zurek mechanism \cite{Kibble1976,Zurek1985,Zurek1996}, the vanishing gap renders adiabatic dynamics impossible, leading to unavoidable excitations that manifest as topological defects in the ordered phase. In the slow-driving regime, these defects are highly coherent and exhibit well-defined momentum-space correlations. During subsequent unitary evolution, these coherent excitations accumulate global phases and, at specific times, they would interfere destructively, causing the Loschmidt echo to vanish--a definitive signature of a dynamical quantum phase transition. Conversely, under high‑frequency driving, the system enters a saturation regime where defect densities become large \cite{Kou2023}, but the excitations are incoherent and lack ordered phase relations. This incoherence, combined with the strong suppression of coherent dynamics by the rapid drive, prevents global quantum interference. Thus, even though a high density of defects is created, no DQPT emerges.
	
	\section{Conclusions}
	\label{Sec: Conclusion}
    This work investigated DQPTs induced by periodic driving in a 1D Ising model, with emphasis on the rate function and the winding number. We found that, unlike sudden quenches, periodic perturbations applied within the same phase can induce a DQPT provided the driving frequency is resonant with the energy-level transition frequency of the system. This resonant effect originates from drive-induced transitions between the system's energy levels and is intimately connected to Floquet topological phases: when the driving frequency lies within the resonant region, the effective Floquet Hamiltonian acquires a nontrivial winding number, signaling the emergence of a topologically distinct nonequilibrium phase. The occurrence of DQPTs precisely coincides with this topological transition. The timescale for the occurrence of a DQPT is governed by the perturbation strength, the critical mode, and its associated energy gap, following the scaling relation $\tau \propto \Delta_{k_c} \lambda'^{-1} \csc k_c$. Furthermore, we examined perturbations that cross the equilibrium critical point, demonstrating that low-frequency drives can trigger DQPTs, whereas high-frequency drives strongly suppress the system's evolution. Compared with conventional quench protocols, periodic perturbations offer a more refined means of controlling DQPTs, thereby providing a new perspective for the study of nonequilibrium dynamics in quantum many-body systems.
	
	\section*{Acknowledgments}
	This work was supported by the Key R\&D Program of Shandong Province, China (Grant No. 2023CXGC010901), and Natural Science Foundation of Shandong Province, China (ZR2024MA018).
	
	\appendix
    \section{Time-dependent perturbation theory} \label{Sec: A}
    Under a transverse field perturbation, the Hamiltonian for mode $k$ can be decomposed into two parts:
	\begin{equation}\label{}
		\hat{H}_k(t)=\hat{H}_k+\hat{H}'_k(t),
	\end{equation}
	where,
	\begin{equation}\label{}
		\hat{H}_k=\begin{pmatrix} \lambda-\cos k &-i\sin k \\ i\sin k & -\lambda+\cos k\end{pmatrix}
	\end{equation}
	denotes the unperturbed Hamiltonian,
	and
	\begin{equation}\label{}
		\hat{H}'_k(t)=\begin{pmatrix} \lambda'\cos\omega t &0 \\ 0 & -\lambda'\cos\omega t \end{pmatrix},
	\end{equation}
	represents the time dependent perturbation.
	The time-evolved state under this perturbation can be represented as
	\begin{equation}
		|\psi_k(t)\rangle = a_k^+(t) e^{-i \varepsilon_{k} t} |\varepsilon^{+}_{k}\rangle + a_k^-(t) e^{i \varepsilon_k t} |\varepsilon^{-}_{k}\rangle.
	\end{equation}
	The coefficient $a^{\pm}_{k}(t)$ admit a perturbative expansion:
	\begin{equation}\label{}
		a^{\pm}_{k}(t)=a_{k}^{\pm(0)}(t)+a_{k}^{\pm(1)}(t)+\cdots,
	\end{equation}
	where $a_{k}^{\pm(0)}(t)$ and $a_{k}^{\pm(1)}(t)$ denote the zeroth- and the first-order perturbation corrections respectively. In the weak-driving regime where $\lambda' < 1$, the higher-order terms can be neglected, and the dynamical behavior of the system can be accurately described by the first-order term.
	The evolution of the perturbation coefficients satisfies the following equations
	\begin{equation}
		\dot{a}_{k}^{\pm(0)}(t) = 0,
	\end{equation}
	and
	\begin{equation}\label{}
		\dot{a}_{k}^{\pm(1)}(t)= -ia_{k}^{\pm(0)}\hat{H'}_{k}^{\pm,\pm}(t)- i a_{k}^{\mp(0)} e^{\pm i2\varepsilon_k t}\hat{H'}_{k}^{\pm,\mp}(t),
	\end{equation}
	where $\hat{H'}_{k}^{\pm,\pm}(t) = \langle\varepsilon_k^\pm| \hat{H}'_k(t)|\varepsilon_k^\pm\rangle=\pm\cos\theta_k\lambda'\cos\omega t$ and $\hat{H'}_{k}^{\pm,\mp}(t) = \langle\varepsilon_k^\pm| \hat{H}'_k(t)|\varepsilon_k^\mp\rangle=\pm i\sin\theta_k\lambda'\cos\omega t$. For the initial ground state $|\varepsilon_k^-\rangle$, the zeroth-order coefficients are
	\begin{equation}
		a^{+(0)}_k = 0, ~~a^{-(0)}_k = 1,
	\end{equation}
	and first-order approximation coefficients are given by
	\begin{equation}
		\begin{split}
			a_k^{+(1)} &= \lambda'\sin\theta_k\biggl[\frac{e^{i(2\varepsilon_k-\omega)t}-1}{i(2\varepsilon_k-\omega)}
			+\frac{e^{i(2\varepsilon_k+\omega)t}-1}{i(2\varepsilon_k+\omega)} \biggr], \\
			a_k^{-(1)} & =i\lambda'\cos\theta_k\frac{\sin\omega t}{\omega}.
		\end{split}
	\end{equation}
	When the driving frequency resonates with the system transition frequency, i.e., $\omega=2\varepsilon_k$, the first-order coefficients simplify to
	\begin{equation}
		\begin{split}
			a_k^{+(1)} &\approx \lambda'\sin\theta_kt, \\
			a_k^{-(1)} & \approx 0,
		\end{split}
	\end{equation}
	where the high-frequency oscillation terms have been neglected. This shows that under resonant driving, the probability amplitude of the excited state grows linearly in time, indicating that the system undergoes a transition into the excited state. The timescale for this transition is inversely proportional to the perturbation strength
	\begin{equation}
		t_k\propto \lambda'^{-1}\csc \theta_k.
	\end{equation}
	According to the definition of $\sin\theta_k=\frac{\sin k}{\varepsilon_k}=\frac{2\sin k}{\Delta_k}$, this transition timescale is jointly determined by energy gap and perturbation strength:
	\begin{equation}
		t_k\propto \Delta_{k}\lambda'^{-1}\csc k.
	\end{equation}

	\section{Floquet Hamiltonian}\label{Sec: B}
	  In the interaction picture, the Hamiltonian takes the form 
    \begin{widetext}
    \begin{equation}\label{}
    \begin{split}
		\hat{H}_k^I(t)=e^{i\hat{H}_kt}\hat{H}'_k(t)e^{-i\hat{H}_kt}
        =\lambda'\cos\omega t\Bigl[\cos\theta_k\tilde\sigma_k^z-\sin\theta_k\Bigl(\tilde\sigma_k^y\cos(2\varepsilon_kt)-\tilde\sigma_k^x\sin(2\varepsilon_kt)\Bigr)\Bigr],
    \end{split}
	\end{equation}
    \end{widetext}
    where $\tilde\sigma_k^z=|\varepsilon_k^+\rangle\langle\varepsilon_k^+|-|\varepsilon_k^-\rangle\langle\varepsilon_k^-|$, $\tilde\sigma_k^x=|\varepsilon_k^+\rangle\langle\varepsilon_k^-|+|\varepsilon_k^-\rangle\langle\varepsilon_k^+|$ and 
    $\tilde\sigma_k^y=-i|\varepsilon_k^+\rangle\langle\varepsilon_k^-|+i|\varepsilon_k^-\rangle\langle\varepsilon_k^+|$.  $|\varepsilon_k^\pm\rangle$ is the eigenvector of $\hat{H}_k$ with the corresponding eigenvalue $\varepsilon_k^\pm$. Ignoring high-frequency oscillations and employing the rotating wave approximation, the interaction-picture Hamiltonian simplifies to 
    \begin{equation}\label{}
    \begin{split}
		\hat{H}_k^I(t)=-\frac{\lambda'\sin\theta_k}{2}\Bigl[\tilde\sigma_k^y\cos(\delta_kt)+\tilde\sigma_k^x\sin(\delta_kt)\Bigr],
    \end{split}
	\end{equation}
    In a rotating frame defined by $\hat{U}_R(t)=e^{-i\delta_k t\tilde\sigma_k^z/2}$, one obtains a time-independent effective Hamiltonian $\hat{H}_k^{F}=\hat{U}_R^\dag(t)\hat{H}_k^I(t)\hat{U}_R(t)+i\dot{\hat{U}}_R^\dag(t)\hat{U}_R(t)$:
    \begin{equation}\label{}
		\hat{H}_k^{F}=-\frac{\delta_k}{2}\tilde\sigma_k^z-\frac{\lambda'\sin\theta_k}{2}\tilde\sigma_k^y.
	\end{equation}
    This Hamiltonian can be interpreted as the Floquet Hamiltonian.


\begin{thebibliography}{}
	\bibitem{Schreiber2015}
	Schreiber M, Hodgman S S, Bordia P, L{\"u}schen H P, Fischer M H, Vosk R, Altman E, Schneider U and Bloch I 2015
	Observation of many-body localization of interacting fermions in a quasirandom optical lattice
	\textit{Science} \textbf{349} 842--845
	
	\bibitem{Jotzu2014}
	Jotzu G, Messer M, Desbuquois R, Lebrat M, Uehlinger T, Greif D and Esslinger T 2014
	Experimental realization of the topological Haldane model with ultracold fermions
	\textit{Nature} \textbf{515} 237--240
	
	\bibitem{Daley2012}
	Daley A J, Pichler H, Schachenmayer J and Zoller P 2012
	Measuring entanglement growth in quench dynamics of bosons in an optical lattice
	\textit{Phys. Rev. Lett.} \textbf{109} 020505
	
	\bibitem{Saffman2010}
	Saffman M, Walker T G and M{\o}lmer K 2010
	Quantum information with Rydberg atoms
	\textit{Rev. Mod. Phys.} \textbf{82} 2313--2363
	
	\bibitem{Klocke2025}
	Klocke K, Simm D, Zhu G-Y, Trebst S and Buchhold M 2025
	Entanglement dynamics in monitored Kitaev circuits: Loop models, symmetry classification, and quantum Lifshitz scaling
	\textit{Phys. Rev. B} \textbf{111} 224301
	
	\bibitem{Cormick2010}
	Cormick C and Paz J P 2010
	Observing different phases for the dynamics of entanglement in an ion trap
	\textit{Phys. Rev. A} \textbf{81} 022306
	
	\bibitem{Knaute2022}
	Knaute J and Hauke P 2022
	Relativistic meson spectra on ion-trap quantum simulators
	\textit{Phys. Rev. A} \textbf{105} 022616
	
	\bibitem{Heyl2015}
	Heyl M 2015
	Scaling and universality at dynamical quantum phase transitions
	\textit{Phys. Rev. Lett.} \textbf{115} 140602
	
	\bibitem{Heyl2013}
	Heyl M, Polkovnikov A and Kehrein S 2013
	Dynamical quantum phase transitions in the transverse-field Ising model
	\textit{Phys. Rev. Lett.} \textbf{110} 135704
	
	\bibitem{Polkovnikov2011}
	Budich J C and Heyl M 2016
	Dynamical topological order parameters far from equilibrium
	\textit{Phys. Rev. B} \textbf{93} 085416
	
	\bibitem{RevModPhys.83.863}
	Polkovnikov A, Sengupta K, Silva A and Vengalattore M 2011
	Colloquium: nonequilibrium dynamics of closed interacting quantum systems
	\textit{Rev. Mod. Phys.} \textbf{83} 863--883
	
	\bibitem{Zurek2005}
	Zurek W H, Dorner U and Zoller P 2005
	Dynamics of a quantum phase transition
	\textit{Phys. Rev. Lett.} \textbf{95} 105701
	
	\bibitem{Heyl2018}
	Heyl M 2018
	Dynamical quantum phase transitions: a review
	\textit{Rep. Prog. Phys.} \textbf{81} 054001
	
	\bibitem{Heyl2019}
	Heyl M 2019
	Dynamical quantum phase transitions: a brief survey
	\textit{Europhys. Lett.} \textbf{125} 26001
	
	\bibitem{Quan2006}
	Quan H T, Song Z, Liu X F, Zanardi P and Sun C P 2006
	Decay of Loschmidt echo enhanced by quantum criticality
	\textit{Phys. Rev. Lett.} \textbf{96} 140604
	
	\bibitem{Jafari2019a}
	Jafari R, Johannesson H, Langari A and Martin-Delgado M A 2019
	Quench dynamics and zero-energy modes: the case of the Creutz model
	\textit{Phys. Rev. B} \textbf{99} 054302
	
	\bibitem{Jafari2017}
	Jafari R and Johannesson H 2017
	Loschmidt echo revivals: critical and noncritical
	\textit{Phys. Rev. Lett.} \textbf{118} 015701
	
	\bibitem{Divakaran2013}
	Divakaran U 2013
	Three-site interacting spin chain in a staggered field: fidelity versus Loschmidt echo
	\textit{Phys. Rev. E} \textbf{88} 052122
	
	\bibitem{Hou2020}
	Hou X-Y, Gao Q-C, Guo H, He Y, Liu T and Chien C-C 2020
	Ubiquity of zeros of the Loschmidt amplitude for mixed states in different physical processes and its implication
	\textit{Phys. Rev. B} \textbf{102} 104305
	
	\bibitem{Najafi2019}
	Najafi K, Rajabpour M A and Viti J 2019
	Return amplitude after a quantum quench in the XY chain
	\textit{J. Stat. Mech.} \textbf{2019} 083102
	
	\bibitem{Yan2020}
	Yan B, Cincio L and Zurek W H 2020
	Information scrambling and Loschmidt echo
	\textit{Phys. Rev. Lett.} \textbf{124} 160603
	
	\bibitem{Najafi2018}
	Najafi K, Rajabpour M A and Viti J 2018
	Light-cone velocities after a global quench in a noninteracting model
	\textit{Phys. Rev. B} \textbf{97} 205103
	
	\bibitem{Mukherjee2019}
	Mukherjee S and Nag T 2019
	Dynamics of decoherence of an entangled pair of qubits locally connected to a one-dimensionaldisordered spin chain
	\textit{J. Stat. Mech.} \textbf{2019} 043108
	
	\bibitem{Andraschko2014}
	Andraschko F and Sirker J 2014
	Dynamical quantum phase transitions and the Loschmidt echo: a transfer matrix approach
	\textit{Phys. Rev. B} \textbf{89} 125120
	
	\bibitem{Karrasch2013}
	Karrasch C and Schuricht D 2013
	Dynamical phase transitions after quenches in nonintegrable models
	\textit{Phys. Rev. B} \textbf{87} 195104
	
	\bibitem{Vajna2014}
	Vajna S and D{\'o}ra B 2014
	Disentangling dynamical phase transitions from equilibrium phase transitions
	\textit{Phys. Rev. B} \textbf{89} 161105
	
	\bibitem{Jafari2021}
	Jafari R and Akbari A 2021
	Floquet dynamical phase transition and entanglement spectrum
	\textit{Phys. Rev. A} \textbf{103} 012204
	
	\bibitem{Vajna2015}
	Vajna S and D{\'o}ra B 2015
	Topological classification of dynamical phase transitions
	\textit{Phys. Rev. B} \textbf{91} 155127
	
	\bibitem{Mendoza-Arenas2022}
	Mendoza-Arenas J J 2022
	Dynamical quantum phase transitions in the one-dimensional extended Fermi-Hubbard model
	\textit{J. Stat. Mech.} \textbf{2022} 043101
	
	\bibitem{Mondal2022}
	Mondal D and Nag T 2022
	Anomaly in the dynamical quantum phase transition in a non-Hermitian system with extended gapless phases
	\textit{Phys. Rev. B} \textbf{106} 054308
	
	\bibitem{DeNicola2021}
	De Nicola S, Michailidis A A and Serbyn M 2021
	Entanglement view of dynamical quantum phase transitions
	\textit{Phys. Rev. Lett.} \textbf{126} 040602
	
	\bibitem{Sedlmayr2018a}
	Sedlmayr N, Jaeger P, Maiti M and Sirker J 2018
	Bulk-boundary correspondence for dynamical phase transitions in one-dimensional topological insulators and superconductors
	\textit{Phys. Rev. B} \textbf{97} 064304
	
	\bibitem{Sedlmayr2018b}
	Sedlmayr N, Fleischhauer M and Sirker J 2018
	Fate of dynamical phase transitions at finite temperatures and in open systems
	\textit{Phys. Rev. B} \textbf{97} 045147
	
	\bibitem{Verga2023}
	Verga A D 2023
	Entanglement dynamics and phase transitions of the Floquet cluster spin chain
	\textit{Phys. Rev. B} \textbf{107} 085116
	
	\bibitem{Khan2023}
	Khan N A, Wang P, Jan M and Gao X 2023
	Anomalous correlation-induced dynamical phase transitions
	\textit{Sci. Rep.} \textbf{13} 9470
	
	\bibitem{Kriel2014}
	Kriel J N, Karrasch C and Kehrein S 2014
	Dynamical quantum phase transitions in the axial next-nearest-neighbor Ising chain
	\textit{Phys. Rev. B} \textbf{90} 125106
	
	\bibitem{Canovi2014}
	Canovi E, Werner P and Eckstein M 2014
	First-order dynamical phase transitions
	\textit{Phys. Rev. Lett.} \textbf{113} 265702
	
	\bibitem{Calabrese2006}
	Calabrese P and Cardy J 2006
	Time dependence of correlation functions following a quantum quench
	\textit{Phys. Rev. Lett.} \textbf{96} 136801
	
	\bibitem{Uhrich2020}
	Uhrich P, Defenu N, Jafari R and Halimeh J C 2020
	Out-of-equilibrium phase diagram of long-range superconductors
	\textit{Phys. Rev. B} \textbf{101} 245148
	
	\bibitem{Yu2021}
	Yu W C, Sacramento P D, Li Y C and Lin H-Q 2021
	Correlations and dynamical quantum phase transitions in an interacting topological insulator
	\textit{Phys. Rev. B} \textbf{104} 085104
	
	\bibitem{Sadrzadeh2021}
	Sadrzadeh M, Jafari R and Langari A 2021
	Dynamical topological quantum phase transitions at criticality
	\textit{Phys. Rev. B} \textbf{103} 144305
	
	\bibitem{Liou2018}
	Liou S-F and Yang K 2018
	Quench dynamics across topological quantum phase transitions
	\textit{Phys. Rev. B} \textbf{97} 235144
	
	\bibitem{Bhattacharjee2018}
	Bhattacharjee S and Dutta A 2018
	Dynamical quantum phase transitions in extended transverse Ising models
	\textit{Phys. Rev. B} \textbf{97} 134306
	
	\bibitem{Yao2025}
	Yao G-Z, Peng L-Y and Gong Q 2025
	Theory of finite-size scaling in dynamical quantum phase transitions
	\textit{Phys. Rev. A} \textbf{112} 042203
	
	\bibitem{Trapin2021}
	Trapin D, Halimeh J C and Heyl M 2021
	Unconventional critical exponents at dynamical quantum phase transitions in a random Ising chain
	\textit{Phys. Rev. B} \textbf{104} 115159
	
	\bibitem{Sharma2014}
	Sharma S, Russomanno A, Santoro G E and Dutta A 2014
	Loschmidt echo and dynamical fidelity in periodically driven quantum systems
	\textit{Europhys. Lett.} \textbf{106} 67003
	
	\bibitem{Bhattacharya2017a}
	Bhattacharya U, Bandyopadhyay S and Dutta A 2017
	Mixed state dynamical quantum phase transitions
	\textit{Phys. Rev. B} \textbf{96} 180303
	
	\bibitem{Eckstein2009}
	Eckstein M, Kollar M and Werner P 2009
	Thermalization after an interaction quench in the Hubbard model
	\textit{Phys. Rev. Lett.} \textbf{103} 056403
	
	\bibitem{Flaeschner2018}
	Fl{\"a}schner N, Vogel D, Tarnowski M, Rem B S, L{\"u}hmann D-S, Heyl M, Budich J C, Mathey L, Sengstock K and Weitenberg C 2018
	Observation of dynamical vortices after quenches in a system with topology
	\textit{Nat. Phys.} \textbf{14} 265--268
	
	\bibitem{Jurcevic2017}
	Jurcevic P, Shen H, Hauke P, Maier C, Brydges T, Hempel C, Lanyon B P, Heyl M, Blatt R and Roos C F 2017
	Direct observation of dynamical quantum phase transitions in an interacting many-body system
	\textit{Phys. Rev. Lett.} \textbf{119} 080501
	
	\bibitem{Guo2019}
	Guo X-Y, Yang C, Zeng Y, Peng Y, Li H-K, Deng H, Jin Y-R, Chen S, Zheng D and Fan H 2019
	Observation of a dynamical quantum phase transition by a superconducting qubit simulation
	\textit{Phys. Rev. Appl.} \textbf{11} 044080
	
	\bibitem{Chen2020a}
	Chen B, Hou X, Zhou F, Qian P, Shen H and Xu N 2020
	Detecting the out-of-time-order correlations of dynamical quantum phase transitions in a solid-state quantum simulator
	\textit{Appl. Phys. Lett.} \textbf{116} 194002
	
	\bibitem{Wang2019}
	Wang K, Qiu X, Xiao L, Zhan X, Bian Z, Yi W and Xue P 2019
	Simulating dynamic quantum phase transitions in photonic quantum walks
	\textit{Phys. Rev. Lett.} \textbf{122} 020501
	
	\bibitem{Xu2020}
	Xu X-Y, Wang Q-Q, Heyl M, Budich J C, Pan W-W, Chen Z, Jan M, Sun K, Xu J-S, Han Y-J, Li C-F and Guo G-C 2020
	Measuring a dynamical topological order parameter in quantum walks
	\textit{Light Sci. Appl.} \textbf{9} 7
	
	\bibitem{Zhang2025}
	Zhang H, Wang K, Xiao L and Xue P 2025
	Self-normal and biorthogonal dynamical quantum phase transitions in non-Hermitian quantum walks
	\textit{Light Sci. Appl.} \textbf{14} 253
	
	\bibitem{PhysRevB.107.L121113}
	Da\c{g} C B, Wang Y, Uhrich P, Na X and Halimeh J C 2023
	Critical slowing down in sudden quench dynamics
	\textit{Phys. Rev. B} \textbf{107} L121113
	
	\bibitem{DeGrandi2010}
	De Grandi C, Gritsev V and Polkovnikov A 2010
	Quench dynamics near a quantum critical point
	\textit{Phys. Rev. B} \textbf{81} 012303
	
	\bibitem{Weiss2018}
	Weiss W, Gerster M, Jaschke D, Silvi P and Montangero S 2018
	Kibble-Zurek scaling of the one-dimensional Bose-Hubbard model at finite temperatures
	\textit{Phys. Rev. A} \textbf{98} 063601
	
	\bibitem{Sim2022}
	Sim K, Chitra R and Molignini P 2022
	Quench dynamics and scaling laws in topological nodal loop semimetals
	\textit{Phys. Rev. B} \textbf{106} 224302
	
	\bibitem{Cao2024}
	Cao K-Y, Hou H-S and Tong P-Q 2024
	Exploring dynamical phase transitions in the XY chain through a linear quench: early and long-term perspectives
	\textit{Phys. Rev. A} \textbf{110} 042209
	
	\bibitem{Poletti2011}
	Poletti D and Kollath C 2011
	Slow quench dynamics of periodically driven quantum gases
	\textit{Phys. Rev. A} \textbf{84} 013615
	
	\bibitem{Chen2020b}
	Chen Z, Cui J-M, Ai M-Z, He R, Huang Y-F, Han Y-J, Li C-F and Guo G-C 2020
	Experimentally detecting dynamical quantum phase transitions in a slowly quenched Ising-chain model
	\textit{Phys. Rev. A} \textbf{102} 042222
	
	\bibitem{Kang2020}
	Kang K-T, Lo C-Y, Yin S and Chen P 2020
	Kibble-Zurek mechanism in a quantum link model
	\textit{Phys. Rev. A} \textbf{101} 023610
	
	\bibitem{Lee2015}
	Lee M, Han S and Choi M-S 2015
	Kibble-Zurek mechanism in a topological phase transition
	\textit{Phys. Rev. B} \textbf{92} 035117
	
	\bibitem{Balducci2023}
	Balducci F, Beau M, Yang J, Gambassi A and del Campo A 2023
	Large deviations beyond the Kibble-Zurek mechanism
	\textit{Phys. Rev. Lett.} \textbf{131} 230401
	
	\bibitem{Xu_2024}
	Xu B-M 2024
	Quantum-coherence-assisted dynamical phase transition in the one-dimensional transverse-field Ising model
	\textit{Commun. Theor. Phys.} \textbf{76} 125104
	
	\bibitem{Quelle2017}
	Quelle A and Smith C M 2017
	Resonances in a periodically driven bosonic system
	\textit{Phys. Rev. E} \textbf{96} 052105
	
	\bibitem{Ansari2025}
	Ansari S, Jafari R, Akbari A and Abdi M 2025
	Scaling and universality at noisy quench dynamical quantum phase transitions
	\textit{Phys. Rev. B} \textbf{112} 054304
	
	\bibitem{Kitagawa2010}
	Kitagawa T, Berg E, Rudner M and Demler E 2010
	Topological characterization of periodically driven quantum systems
	\textit{Phys. Rev. B} \textbf{82} 235114
	
	\bibitem{Yang2021}
	Yang X-Q and Cai Z 2021
	Dynamical transitions and critical behavior between discrete time crystal phases
	\textit{Phys. Rev. Lett.} \textbf{126} 020602
	
	\bibitem{Brown2021}
	Brown K, Bland T, Comaron P and Proukakis N P 2021
	Periodic quenches across the Berezinskii-Kosterlitz-Thouless phase transition
	\textit{Phys. Rev. Res.} \textbf{3} 013097
	
	\bibitem{Bordia2017}
	Bordia P, L{\"u}schen H, Schneider U, Knap M and Bloch I 2017
	Periodically driving a many-body localized quantum system
	\textit{Nat. Phys.} \textbf{13} 460--464
	
	\bibitem{Murali2025}
	Murali A, Guha Sarkar T and Bandyopadhyay J N 2025
	Adiabatic modulation of driving protocols in periodically driven quantum systems
	\textit{Phys. Rev. A} \textbf{111} 022430
	
	\bibitem{Nandy2018}
	Nandy S, Sen A and Sen D 2018
	Steady states of a quasiperiodically driven integrable system
	\textit{Phys. Rev. B} \textbf{98} 245144
	
	\bibitem{Puskarov2016}
	Puskarov T and Schuricht D 2016
	Time evolution during and after finite-time quantum quenches in the transverse-field Ising chain
	\textit{SciPost Phys.} \textbf{1} 003
	
	\bibitem{Cao2022}
	Cao K-Y, Zhong M and Tong P-Q 2022
	Dynamical quantum phase transition in periodic quantum Ising chains
	\textit{J. Phys. A: Math. Theor.} \textbf{55} 365001

    \bibitem{Yang2019}
     Yang K, Zhou L W, Ma W C, Kong X, Wang P F, Qin X, Rong X, Wang Y, Shi F Z, Gong J B and Du J F 2019
     Floquet dynamical quantum phase transitions
     \textit{Phys. Rev. B} \textbf{100} 085308

	\bibitem{Bhattacharya2017b}
	Bhattacharya U and Dutta A 2017
	Emergent topology and dynamical quantum phase transitions in two-dimensional closed quantum systems
	\textit{Phys. Rev. B} \textbf{96} 014302.

    \bibitem{Bukov2015}
    Bukov M, D'Alessio L, Polkovnikov A 2015
    Universal high-frequency behavior of periodically driven systems: from dynamical stabilization to Floquet engineering
    \textit{Adv. Phys.} \textbf{64}, 139--226 (2015).

    \bibitem{Rudner2013}
    Rudner M S, Lindner N H, Berg E, Levin M 2013
    Anomalous Edge States and the Bulk-Edge Correspondence for Periodically Driven Two-Dimensional Systems
    \textit{Phys. Rev. X} \textbf{3} 031005.

    \bibitem{Lindner2011}
    Lindner N H, Refael G, Galitski V 2011
    Floquet topological insulator in semiconductor quantum wells
    \textit{Nat. Phys.} \textbf{7}, 490--495 (2011)

    \bibitem{Kibble1976}
    Kibble T W B 1976
    Topology of cosmic domains and strings
    \textit{J. Phys. A: Math. Gen.} \textbf{9}, 1387 (1976).

    \bibitem{Zurek1985}
    Zurek W H 1985
    Cosmological experiments in superfluid helium?
    \textit{Nature} \textbf{317} 505--508
    
    \bibitem{Zurek1996}
    Zurek W H 1996
    Cosmological experiments in condensed matter systems
    \textit{Phys. Rep.} \textbf{276} 177-221.

    \bibitem{Kou2023}
    Kou H-C and Li P 2023
    Varying quench dynamics in the transverse Ising chain: The Kibble-Zurek, saturated, and presaturated regimes
    \textit{Phys. Rev. B} \textbf{108} 214307.

	\end{thebibliography}
\end{document}